\documentstyle[amssymb,epsfig,12pt]{article}   
\newlength{\dinwidth}                     
\newlength{\dinmargin}                      
\begin{document}
%
%%%%%%%%%%%%%%%%%%%%%%%%%%%%
% DUMMY TITLE PAGE FOR HEPPH
%%%%%%%%%%%%%%%%%%%%%%%%%%%%
\thispagestyle{empty}
\begin{flushright}
\large
DTP/99/86 \\
ITP-SB-99-38 \\
July 1999
\end{flushright}
\vspace{1.2cm}

\renewcommand{\thefootnote}{\fnsymbol{footnote}}
\setcounter{footnote}{1}
\begin{center}  \begin{Large} \begin{bf}
\hbox to\textwidth{\hss
{Towards the Parton Densities of Polarized Photons at 
HERA\footnote{Talk presented by M.~Stratmann at the workshop on
`Polarized Protons at High Energies - Accelerator Challenges and
Physics Opportunities', DESY, Germany, May 1999.}} \hss}
\end{bf}  \end{Large}

\vspace*{1.8cm}
{\Large M.~Stratmann}

\vspace*{4mm}

{Department of Physics, University of Durham,}\\

\vspace*{2mm}
{Durham DH1 3LE, England}\\

\vspace*{7mm}

{and}\\

\vspace*{0.8cm}

{\Large W.~Vogelsang}

\vspace*{4mm}

{C.N.\ Yang Institute for Theoretical Physics, State University of
New York at Stony Brook,}

\vspace*{2mm}
{NY-11794, USA}

\vspace*{2.cm}

\normalsize
{\large \bf Abstract}
\end{center}

\noindent
Di-jet photoproduction in polarized $ep$ collisions at HERA is studied 
as a possible tool to determine the parton content of circularly polarized 
photons. The concept of the `effective parton density' approximation 
is extended to the spin-dependent case.
\setcounter{page}{0}
\renewcommand{\thefootnote}{\arabic{footnote}}
\setcounter{footnote}{0}
\normalsize
\newpage

% Start of document
% -----------------
%%%\begin{document}
\vspace*{10mm}
\begin{center}  \begin{Large} \begin{bf}
\hbox to\textwidth{\hss
{Towards the Parton Densities of Polarized Photons at HERA} \hss}
\end{bf}  \end{Large}
\vspace*{5mm}
\begin{large}
M.\ Stratmann$^a$ and W.\ Vogelsang$^b$
\end{large}
\end{center}
$^a$ Department of Physics, University of Durham, Durham DH1 3LE, England\\
$^b$ C.N.\ Yang Institute for Theoretical Physics, State University of 
New York at Stony Brook, NY-11794, USA
\begin{quotation}
\noindent
{\bf Abstract:}
Di-jet photoproduction in polarized $ep$ collisions at HERA is studied 
as a possible tool to determine the parton content of circularly polarized 
photons. The concept of the `effective parton density' approximation 
is extended to the spin-dependent case.
\end{quotation}
\renewcommand{\thefootnote}{\arabic{footnote}}
\setcounter{footnote}{0}

%%%%%%%%%%%%%%%%%%%%%%%%%%%%%%%%%%%%%%%%%%%%
\section{Introduction and General Framework}
%%%%%%%%%%%%%%%%%%%%%%%%%%%%%%%%%%%%%%%%%%%%
%
Previous studies \cite{ref:heraresults1,ref:heraresults2} have exposed
the photoproduction of (di-)jets in longitudinally polarized 
$ep$ collisions at HERA as a very promising {\em and} feasible 
tool to measure the parton densities of circularly polarized photons in
`resolved'-photon processes. It should be stressed 
that these photonic parton distributions, defined as
\begin{equation}
\label{eq:pdfdef}
\Delta f^{\gamma}(x,Q^2) \equiv f_+^{\gamma_{+}}(x,Q^2) -  
f_-^{\gamma_{+}}(x,Q^2)\;\;,
\end{equation}
where $f_+^{\gamma_{+}}$ $(f_-^{\gamma_{+}})$ 
denotes the density of a parton $f$ with helicity `+' (`$-$') 
in a photon with helicity `+', are completely unmeasured so far.
The $\Delta f^{\gamma}$ contain information different
from that included in the unpolarized $f^{\gamma}$ [defined
by taking the sum in (\ref{eq:pdfdef})], and their measurement is
indispensable for a thorough understanding of the 
partonic structure of the photon.
As in \cite{ref:heraresults1,ref:heraresults2} we will exploit the 
predictions of two very different models for the $\Delta f^{\gamma}$ 
\cite{ref:gsv}, and study the sensitivity of di-jet production to these 
unknown quantities. In the first case (`maximal scenario') 
we saturate the positivity 
bound $|\Delta f^{\gamma}(x,Q^2)| \le f^{\gamma}(x,Q^2)$
at a low input scale $\mu\simeq 0.6\,{\rm{GeV}}$, using the
unpolarized GRV densities $f^{\gamma}$ \cite{ref:grvphot}.
The other extreme input (`minimal scenario') is defined by
a vanishing hadronic input at the same scale $\mu$.
We limit ourselves to leading order (LO) QCD, 
which is entirely sufficient for our purposes; however
both scenarios can be straightforwardly extended to the
next-to-leading order (NLO) of QCD \cite{ref:nloletter}.

The generic expression for polarized photoproduction of two jets 
with laboratory system rapidities $\eta_1$, $\eta_2$ reads in LO
\begin{equation} 
\label{eq:wq2jet}
\frac{d^3 \Delta \sigma}{dp_T d\eta_1 d\eta_2} = 2 p_T
\sum_{f^e,f^p} x_e \Delta f^e (x_e,\mu_f^2) x_p \Delta f^p (x_p,\mu_f^2)
\frac{d\Delta \hat{\sigma}}{d\hat{t}} \; ,
\end{equation}
where $p_T$ is the transverse momentum of one of the two jets (which balance
each other in LO),
$x_e \equiv p_T/(2 E_e) \left( e^{-\eta_1} + e^{-\eta_2} \right)$, and
$x_p \equiv p_T/(2 E_p) \left( e^{\eta_1} + e^{\eta_2} \right)$.
The $\Delta f^p$ in (\ref{eq:wq2jet}) denote the spin-dependent parton 
densities of the proton, and\footnote{The direct (`unresolved') 
photon contribution to (\ref{eq:wq2jet}) is obtained by setting
$\Delta f^{\gamma} (x_{\gamma},\mu_f^2) \equiv \delta (1-x_{\gamma})$
in (\ref{eq:elec}).}
\begin{equation}  
\label{eq:elec}
\Delta f^e (x_e,\mu_f^2) = \int_{x_e}^1 \frac{dy}{y} \Delta P_{\gamma/e} (y)
\Delta f^{\gamma} (x_{\gamma}=\frac{x_e}{y},\mu_f^2) \;\;,
\end{equation}
where $\Delta P_{\gamma /e}$ is the polarized `equivalent-photon'
spectrum for which we will use\footnote{Very recently
the non-logarithmic corrections to (\ref{eq:weiz}) have been calculated
in \cite{ref:nlojets1}. They typically lead to an ${\cal{O}}(10\,\%)$
correction which, however, cancels to a large extent in the experimentally
relevant spin asymmetry $\Delta \sigma/\sigma$, and thus can be safely
neglected here.}
\begin{equation}  
\label{eq:weiz}
\Delta P_{\gamma/e} (y) = \frac{\alpha_{em}}{2\pi} \left[
\frac{1-(1-y)^2}{y} \right] \ln \frac{Q^2_{\mathrm{max}} (1-y)}{m_e^2 y^2} \; ,
\end{equation}
with the electron mass $m_e$ and $Q^2_{\mathrm{max}}=4\,\mathrm{GeV}^2$.
Needless to say that the unpolarized LO jet cross section
$d^3\sigma$ is obtained by using the corresponding 
unpolarized quantities in (\ref{eq:wq2jet})-(\ref{eq:weiz}).
The appropriate LO $2\rightarrow 2$ partonic cross sections
$d(\Delta)\hat{\sigma}$ in (\ref{eq:wq2jet}) for the direct 
$(\gamma b\rightarrow cd)$ and resolved $(ab\rightarrow cd)$ cases
can be found, for instance, in \cite{ref:gastmans}. 

The key feature of {\em di}-jet production is that a measurement of both
jet rapidities allows for fully reconstructing the kinematics of the 
underlying hard subprocess and thus for determining the 
variable \cite{ref:jet2ph}
$x_{\gamma}^{\mathrm{OBS}} = \sum_{jets} p_T^{jet} e^{-\eta^{jet}}/
(2 y E_e)$, which to LO equals $x_{\gamma}=x_e/y$, 
with $y$ being the fraction of the electron's energy taken by 
the photon. In this way it becomes possible to experimentally 
suppress the direct contribution by introducing some suitable 
cut\footnote{To achieve a similar `separation' for single-inclusive 
jet production one has to `look' into different rapidity directions
since the direct (resolved) contribution dominates in the electron
(proton) direction \cite{ref:heraresults1}.}
$x_{\gamma}^{\mathrm{OBS}} \leq 0.75$ \cite{ref:jeff},
or by scanning different $x_{\gamma}^{\mathrm{OBS}}$ bins.
In \cite{ref:heraresults1,ref:heraresults2} the usefulness of 
this method was demonstrated also for the polarized case. 
In addition it was shown \cite{ref:heraresults2} that
the LO QCD parton level calculations nicely agree with `real' jet
production processes including initial and final state QCD radiation
as well as non-perturbative effects such as hadronization, as 
modeled using the {\tt SPHINX} Monte-Carlo \cite{ref:sphinx}.
These results were all very encouraging; however it was not studied how
one can actually unfold the $\Delta f^{\gamma}$ from such
a measurement. This question will be addressed here.
In this context the concept of `effective parton densities',
developed many years ago \cite{ref:effpdf} and recently revived 
\cite{ref:effpdfunp}, proves to be a useful tool.
We will first recall the basic idea behind this approximation and
subsequently discuss its extension to the spin-dependent case.
 
%%%%%%%%%%%%%%%%%%%%%%%%%%%%%%%%%%%%%%%%%%%%%%
\section{`Effective' Parton Densities Revisited}
%%%%%%%%%%%%%%%%%%%%%%%%%%%%%%%%%%%%%%%%%%%%%%
%
Obviously it would be a very involved task to unfold 
the $\Delta f^{\gamma}$ from a jet-measurement since many 
subprocesses and combinations of parton densities contribute
to the cross section (\ref{eq:wq2jet}). Some handy but still accurate
approximation for (\ref{eq:wq2jet}) is certainly required to facilitate
this job. 

In the unpolarized case a useful approximation procedure was
developed in \cite{ref:effpdf}. It was observed that 
the ratios of the dominant, properly symmetrized, LO subprocesses 
are roughly independent of the c.m.s\ partonic
scattering angle $\Theta$ and, most importantly, that for 
$\cos\Theta=\pm 1$ {\em all} ratios tend to the {\em same} value
determined by the color factors $C_A$ and $C_F$: 
\begin{equation}
\label{eq:unpratios}
\frac{\hat{\sigma}_{qq'}}{\hat{\sigma}_{qg}}\Bigg|_{\cos\Theta=\pm 1} = 
\frac{\hat{\sigma}_{qq}}{\hat{\sigma}_{qg}}\Bigg|_{\cos\Theta=\pm 1} = 
\frac{\hat{\sigma}_{qg}}{\hat{\sigma}_{gg}}\Bigg|_{\cos\Theta=\pm 1} = 
\frac{C_F}{C_A}=\frac{4}{9}\;\;.
\end{equation}
Making use of (\ref{eq:unpratios}) for all values of $\Theta$ 
and introducing the `effective' parton density combinations
%
%%%%%%%%%%%%
% FIGURE 1 %
%%%%%%%%%%%%
\begin{figure}[tbh]
\begin{center}
\vspace*{-1.4cm}
\epsfig{file=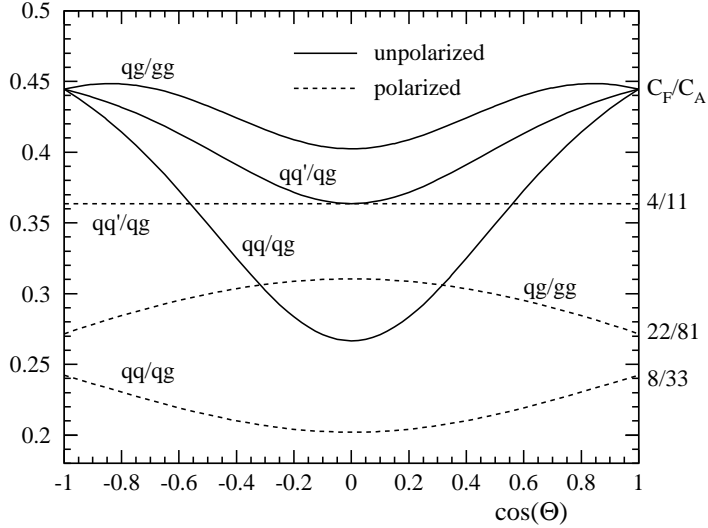,width=10cm}
\end{center}
\vspace*{-6.5cm}
\caption{\sf Ratios of unpolarized and polarized LO 
$2\rightarrow 2$ partonic jet cross sections.}
\vspace*{-0.2cm}
\end{figure}
\begin{equation}
\label{eq:unpeffpdf}
f_{\mathrm{eff}}^{(p,\gamma)} \equiv 
\sum_q [q^{(p,\gamma)}+\bar{q}^{(p,\gamma)}]+\frac{9}{4}g^{(p,\gamma)} \; ,
\end{equation}
the jet cross section factorizes into these densities times a 
{\em single} subprocess cross section (cf.\ Eq.~(\ref{eq:factwq}) below).
The ratios of the parton cross sections are depicted in Fig.~1, and,
although they considerably deviate from $4/9$ for
$\cos \Theta\neq \pm 1$, the approximation works amazingly well at a level
of about ${\cal{O}}(10\%)$ accuracy. 

Unfortunately this approximation 
has no straightforward extension to the spin-dependent case
as is obvious from Fig.~1. The ratios of the LO polarized subprocess
cross sections obey
\begin{equation}
\label{eq:polratios}
\frac{\Delta\hat{\sigma}_{qq'}}{\Delta\hat{\sigma}_{qg}}\Bigg|_{\cos\Theta=
\pm 1} = \frac{4}{11}\,\,\,\,,\,\,\,\,
\frac{\Delta\hat{\sigma}_{qq}}{\Delta\hat{\sigma}_{qg}}\Bigg|_{\cos\Theta=
\pm 1} =  \frac{8}{33}\,\,\,\,,\,\,\,\,
\frac{\Delta\hat{\sigma}_{qg}}{\Delta\hat{\sigma}_{gg}}\Bigg|_{\cos\Theta=
\pm 1} =  \frac{22}{81} 
\end{equation}
rather than approaching a common value for $\cos \Theta=\pm 1$, and,
consequently, the factorization as outlined above is bound to fail. 
However, one also notices that all spin-dependent ratios in
Fig.~1 are more flattish w.r.t\ $\cos \Theta$ than in the 
unpolarized case, and $qq'/qg=4/11$ is {\em exact} for all values of
$\cos \Theta$. It turns out that by approximating all
ratios by $4/11$ and introducing
\begin{equation}
\label{eq:poleffpdf}
\Delta f_{\mathrm{eff}}^{(p,\gamma)} \equiv 
\sum_q [\Delta q^{(p,\gamma)}+ \Delta \bar{q}^{(p,\gamma)}]+
\frac{11}{4} \Delta g^{(p,\gamma)}  \; ,
\end{equation}
the effective parton density approximation works remarkably well also in
this case, and (\ref{eq:wq2jet}) factorizes, e.g., for the resolved 
contribution, schematically into
\begin{equation}
\label{eq:factwq}
d\Delta \sigma^{\mathrm{2-jet}} \simeq 
\int \Delta f_{\mathrm{eff}}^{\gamma} \;
     \Delta f_{\mathrm{eff}}^{p}\;
     d\Delta \hat{\sigma}_{qq'\rightarrow qq'}\,\,.
\end{equation}
%
%%%%%%%%%%%%
% FIGURE 2 %
%%%%%%%%%%%%
\begin{figure}[tbh]
\begin{center}
\vspace*{-1.6cm}
\epsfig{file=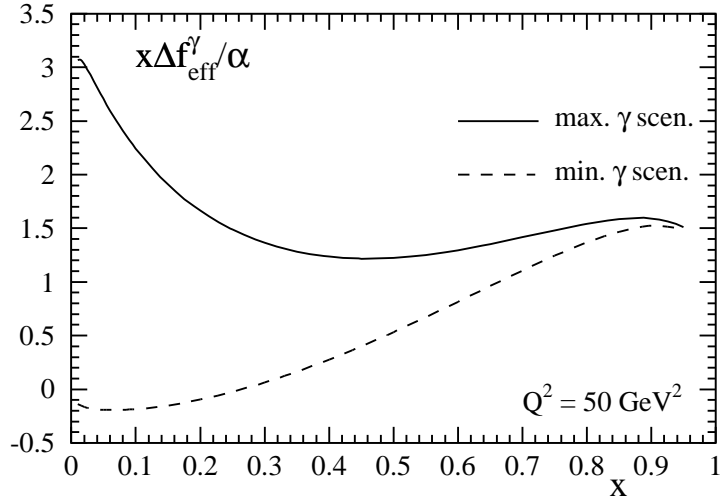,width=10cm}
\end{center}
\vspace*{-6.8cm}
\caption{\sf The effective parton density $\Delta f_{\mathrm{eff}}^{\gamma}$
as defined in (\ref{eq:poleffpdf}) for the two extreme scenarios 
specified in the text, at a scale $Q$ typically probed in jet production.}
\vspace*{-0.2cm}
\end{figure}
For all relevant purposes the approximated and the exact polarized
LO di-jet cross sections agree within $\leq 5\%$, even better
than what is achieved in the unpolarized case.
Figure 2 shows the polarized effective photon density according to
(\ref{eq:poleffpdf}) for the two extreme scenarios specified above
at a scale relevant for the production of jets with 
$p_T$ values of about $5-10\,\mathrm{GeV}$. 

%%%%%%%%%%%%%%%%%%%%%%%%%%%%%%%%%
\section{Results and Conclusions}
%%%%%%%%%%%%%%%%%%%%%%%%%%%%%%%%%
%
Figure 3 shows the experimentally relevant di-jet spin asymmetry
$A^{\mathrm{2-jet}}\equiv d\Delta\sigma/d\sigma$,
for three different bins in $x_{\gamma}$, using
similar cuts as in \cite{ref:effpdfunp}:
the difference of the jet pseudorapidities is required to be
$|\Delta \eta^{\mathrm{jets}}|<1$, for the
average rapidity we demand $0<(\eta_1+\eta_2)/2<1$, and $0.2<y<0.83$.
The factorization scale $\mu_F$ in (\ref{eq:wq2jet}) was chosen to
be equal to $p_T$, but the asymmetry is largely independent of that
choice. Very recently, the complete NLO QCD corrections to polarized
jet-(photo)production have become available 
\cite{ref:nlojets2,ref:nlojets1}.
They lead to an improved scale dependence of the cross sections. 
Moderate NLO corrections for the asymmetry were found for the 
single-inclusive case \cite{ref:nlojets1}; 
similar results should be expected also for di-jet production.

As can be inferred from Fig.~3, the effective parton density 
approximation works very well. It is only for $0.4\le x_{\gamma}\le 0.75$ 
and large $p_T$ that the deviations from the exact results 
become more pronounced.  
%
%%%%%%%%%%%%
% FIGURE 3 %
%%%%%%%%%%%%
\begin{figure}[tbh]
\begin{center}
\vspace*{-1.6cm}
\epsfig{file=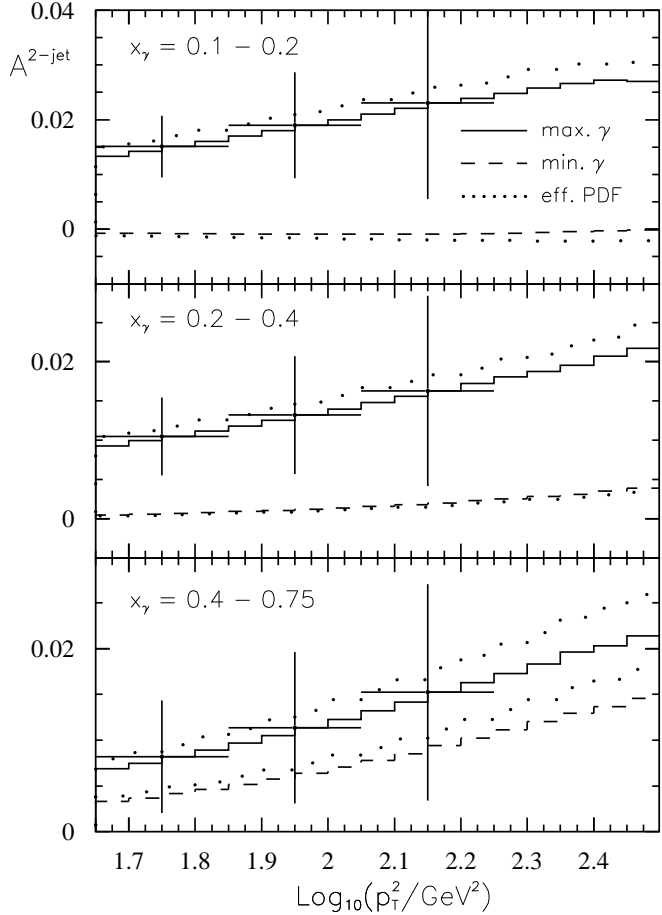,width=10.0cm}
\end{center}
\vspace*{-1.1cm}
\caption{\sf Predictions for $A^{\mathrm{2-jet}}$ for three different
bins in $x_{\gamma}$, using the two scenarios for
$\Delta f^{\gamma}$ as described in the text and the
LO GRSV `standard' distributions \cite{ref:grsv} for $\Delta f^p$. 
Also shown are the results using the effective parton density
approximation (dotted lines) as outlined in Sec.~2 
and the expected statistical errors for such a measurement (see text).}
\vspace*{-0.2cm}
\end{figure}
Also shown in Fig.~3 is the expected statistical accuracy 
for such measurements, assuming three bins in $p_T$ 
for each $x_{\gamma}$ bin, an integrated
luminosity of $200\,\mathrm{pb}^{-1}$, and $70\%$ beam polarizations.
Given these error bars the prospects for distinguishing between
different scenarios for $\Delta f^{\gamma}_{\mathrm{eff}}$ 
are rather promising {\em provided} the proton 
densities $\Delta f^{p}_{\mathrm{eff}}$, also entering 
(\ref{eq:factwq}), are known fairly well, which is clearly not the case yet. 
However, our ignorance of the $\Delta f^{p}$ 
will be vastly reduced by the upcoming polarized $pp$ collider 
RHIC and ongoing efforts in the fixed target sector by HERMES and (soon)
by COMPASS.
It should be kept in mind that so far {\em nothing at all} is known about
the $\Delta f^{\gamma}$, and even to establish the very existence
of a resolved component also in the spin-dependent case would be
an important step forward. 

%%%%%%%%%%%%%%%%%%%%%%%%%%%

%
\end{document}